\documentclass{article}
\usepackage{amssymb}
\usepackage{amsfonts}
\usepackage{amsmath}

\input{tcilatex}
\begin{document}

\title{\textbf{Scalar Casimir effect between two concentric D-dimensional
spheres}}
\author{Mustafa \"{O}zcan\thanks{%
e-mail: ozcanm@trakya.edu.tr and ozcanm99@gmail.com } \\
Department of Physics, Trakya University 22030 Edirne, Turkey}
\maketitle

\begin{abstract}
The Casimir energy for a massless scalar field between the closely spaced
two concentric $D-$dimensional $\left( \text{for }D>3\right) $ spheres is
calculated by using the mode summation with contour integration in the
complex plane of eigenfrequencies and the generalized Abel-Plana formula for
evenly spaced eigenfrequency at large argument. The sign of the Casimir
energy between closely spaced two concentric $D$-dimensional spheres for a
massless scalar field satisfying the Dirichlet boundary conditions is
strictly negative. The Casimir energy between $D-1$ dimensional surfaces
close to each other is regarded as interesting both by itself and as the key
to describing of stability of the attractive Casimir force.

PACS number(s): 03.70.+k, 11.10.Kk, 11.10.Gh, 03.65.Ge
\end{abstract}

\section{INTRODUCTION}

\qquad The measurable consequences of the macroscopic phenomena in the
quantum theory of fields is Casimir effect. This effect due to the vacuum
polarization of the quantized field was originally derived by H. B. G.
Casimir \cite{Cas}. He calculated the negative renormalized quantum vacuum
energy for the electromagnetic field bounded by two uncharged parallel
plates. He concluded that there must exist an attractive force between the
plates. The attractive force: $F=-\frac{\pi ^{2}\hbar c}{240a^{4}}$ which is
expressed in terms of the Planck constant $\hbar ,$ the velocity of light $c$%
, and the distance between the plates $a$, should act on unit area of two
uncharged conducting plane parallel plates in vacuum. This attractive force
was confirmed experimentally \cite{Spaar,Lam,klim}. Both theoretical and
experimental studies of the Casimir effect can provide great insight into
understanding the nature of the quantum vacuum \cite%
{Gunter,Mosbook,Miltonbook,Mos1,Bordagbook}

\qquad At first time, having found the negative energy for uncharged
conducting parallel plates due to Casimir effect, the hope was what the same
would appear for the spherical geometry. This expectation was shattered by
Boyer \cite{Boyer}. Boyer first showed Casimir energy for the spherical
shell is positive. The sign of the positive Casimir energy produces the
repulsive force. Boyer's result has been later confirmed by using different
regularization techniques \cite{Milton1,Nest1,Hag}. Nowadays, the nature of
the Casimir energy is the strong dependence on the geometry of the
spacetime, the dimension of the spacetime and on the boundary condition
imposed \cite{Miltonbook,Mozcan3,Mozcan2}. Recently, the Casimir energy
between two concentric spheres and cylinders in $D=3$ dimension have been
considered by using the different regularization methods \cite%
{Brevik,Brevik1,Mozcan1,MPRD,Teo1,Teo11,Miltao,IHD,Jiang,Tatur,Mazzitelli,Maz1,Funaro,emig,emig1,emig2}
for the scalar and electromagnetic fields. Moreover, the investigation of
the dimensional dependence of the Casimir energy is of interest. The Casimir
energy in the higher dimensional spacetime has a long history. One of the
pioneering works is that the Casimir energy for a massless scalar field and
a massless vector field in a $D-$dimensional rectangular cavity were derived
by Ambjorn and Wolfram \cite{JanAmbjorn}. Afterwards, the scalar and
electromagnetic Casimir energy for a $D-$dimensional sphere were calculated
using Green's functions method and zeta function regularization \cite%
{MiltonD,MiltonD1,ElizaldeD,Fucci,Fucci1}, and the Casimir interaction
between two concentric spheres in $D$-dimensions for the scalar fields with
Robin boundary condition and the electromagnetic fields at finite
temperature have been studied \cite{Aram,Teo2}. In particular today,
physical theories and models with extra dimensions are active areas of
research, for example String theory and Brane world models.

\qquad As matter of fact, the sign and magnitude of the Casimir energy may
strongly depend on (a) the spacetime dimensionality, (b) the type of the
boundary conditions, (c) type of the fields, (d) the lengths between the
surfaces is critically dependent on their nanometre-scale shape, (e) the
curved spacetime background, (f) compactness spacetime, and (g) the finite
temperature \cite{Miltonbook,Li}. In this work, the consequences of (c), (e)
and (g) will not be considered, and we consider the Casimir energy between
closely spaced two concentric $D-$dimensional spheres for a massless scalar
field satisfying the Dirichlet boundary conditions. From a mathematical
point of view, a simple method for calculating the Casimir energy between
closely spaced two concentric $D-$dimensional spheres for a massless scalar
field is developed which is based on a direct mode summation with the
contour integration in a complex plane of eigenfrequencies and using the
generalized Abel-Plana sum formula for evenly spaced eigenfrequency at large
argument. One of the motivations of our work for us to perform the Casimir
energy calculation for a massless scalar field between $D-1$ dimensional
surfaces close to each other is that the physics in higher dimensional
spacetime have become a trend since the existence of the extra dimensions
allows the solving of the some fundamental problems in physics as the
hirarchy problem.

\qquad The organization of this paper is as follows. In section 2, The
Casimir energy of a massless scalar field subjected to the Dirichlet
boundary conditions on between closely spaced two concentric $D-$dimensional
spheres is calculated without any approximation techniques. Concluding
remarks and discussion of the Casimir energy for a massless scalar field in
an annular region of $D-$dimensional geometry is presented in section 3.

\qquad The units are such that $\hslash =c=1$.

\qquad

\bigskip

\bigskip

\section{\protect\bigskip CASIMIR ENERGY}

\qquad We begin with a massless scalar field considered in a $D-$dimensional
spherical geometry, where the metric is given by

\begin{eqnarray}
ds^{2} &=&dt^{2}-d\sigma ^{2},\text{\ \ \ where}  \notag \\
d\sigma ^{2} &=&dx_{1}^{2}+dx_{2}^{2}+dx_{3}^{2}+.......+dx_{D}^{2}.
\end{eqnarray}%
In spherical coordinates

\begin{eqnarray}
x_{1} &=&r\ \sin \theta _{1}\ \sin \theta _{2}\ \sin \theta _{3}.......\sin
\theta _{D-2}\ \cos \phi  \notag \\
x_{2} &=&r\ \sin \theta _{1}\ \sin \theta _{2}\ \sin \theta _{3}.......\sin
\theta _{D-2}\ \sin \phi  \notag \\
&&.  \notag \\
&&.  \notag \\
&&.  \notag \\
x_{D-1} &=&r\ \sin \theta _{1}\ \cos \theta _{2}  \notag \\
x_{D} &=&r\ \cos \theta _{1}\ .
\end{eqnarray}%
Where $0\leq \theta _{j}\leq \pi $, \ $j=1,2,3,....,D-2$ and $0\leq \phi
\leq 2\pi $. $x_{1}^{2}+x_{2}^{2}+x_{3}^{2}+....+x_{D}^{2}=r^{2}$ is defined
by $D$ dimensional hypersphere. The metric becomes

\begin{equation}
ds^{2}=dt^{2}-\left[ dr^{2}+r^{2}d\theta _{1}^{2}+r^{2}\sin ^{2}\theta
_{1}d\theta _{2}^{2}+....+r^{2}\sin ^{2}\theta _{1}\sin ^{2}\theta
_{2}......\sin ^{2}\theta _{D-2}d\phi ^{2}\right] .
\end{equation}

Massless scalar field satisfies the Klein Gordon equation in this geometry
is given by

\begin{equation}
\square \Psi \left( t,r,\vartheta ,\phi \right) =0\ \ \left( \text{where }%
\vartheta =\theta _{1},\theta _{2},.....,\theta _{D-2}\right) .
\end{equation}%
Here $\square $ is the D'Alembertian operator associated with the metric
given by the line element in Eq. (3). Solution of Eq. (4) could be easily
found by using the method of separation of variables and is written as $%
\left( \text{for }D\geq 4\right) $

\begin{eqnarray}
\Psi \left( t,r,\vartheta ,\phi \right) &=&\dsum\limits_{\left\{ \lambda
\right\} }\ e^{-i\omega t}\left[ A_{k}r^{-\frac{\left( D-2\right) }{2}%
}J_{\nu }\left( \omega r\right) +B_{k}r^{-\frac{\left( D-2\right) }{2}%
}N_{\nu }\left( \omega r\right) \right]  \notag \\
&&  \notag \\
&&\left\{ \dprod\limits_{\mu =1}^{D-3}\ \sin ^{M_{D-\mu -2}}\theta _{\mu }\
C_{M_{D-\mu -1}-M_{D-\mu -2}}^{M_{D-\mu -2}+\frac{\left( D-\mu -1\right) }{2}%
}\left( \cos \theta _{\mu }\right) \right\} Y_{\ell m}\left( \theta
_{D-2},\phi \right) \ .  \notag \\
&&
\end{eqnarray}%
Where $\nu =k+\frac{\left( D-2\right) }{2}$ and $\left\{ \lambda \right\} $
refers to $m,\ell \left( =M_{1}\right) ,M_{2},M_{3},M_{4},.....,M_{D-3},$
and $k\left( =M_{D-2}\right) .$ $D$ is the number of dimension of sphere. $%
J_{\nu }\left( \omega r\right) $ and $N_{\nu }\left( \omega r\right) $ are
Bessel functions first and second kind, respectively. And the function $%
C_{M_{D-\mu -1}-M_{D-\mu -2}}^{M_{D-\mu -2}+\frac{\left( D-\mu -1\right) }{2}%
}\left( \cos \theta _{\mu }\right) $ corresponds to the Gegenbauer or
ultraspherical polynomials. $Y_{\ell m}\left( \theta _{D-2},\phi \right) $
are the spherical harmonics. We note that the spherical boundary condition
at $\Psi \left( t,r=a,\vartheta ,\phi \right) =0$ and $\Psi \left(
t,r=b,\vartheta ,\phi \right) =0$ ($a$ ($b$)is inner (outer) radius of $D$%
-dimensional sphere) has not imposed on Eq. (5) yet. Hence, $\omega $ still
remains a continous parameter, while $m,\ell \left( =M_{1}\right)
,M_{2},M_{3},....,k\left( =M_{D-2}\right) $ take the values%
\begin{eqnarray}
m &=&-\ell ,-\ell +1,.....,0,1,2,.....,\ell -1,\ell \ \ ,  \notag \\
\ell \left( =M_{1}\right) &=&0,1,2,3,4,.........,M_{2}  \notag \\
M_{2} &=&0,1,2,3,4,.........,M_{3}  \notag \\
M_{3} &=&0,1,2,3,4,.........,M_{4}  \notag \\
&&.  \notag \\
&&.  \notag \\
&&.  \notag \\
M_{D-4} &=&0,1,2,3,4,.........,M_{D-3}  \notag \\
M_{D-3} &=&0,1,2,3,4,.........,M_{D-2}  \notag \\
k\left( =M_{D-2}\right) &=&0,1,2,3,4,.........\ \ .
\end{eqnarray}

We now impose the boundary conditions for $D-$dimensional spherical geometry
i.e.,

\begin{equation}
\Psi _{\omega \ell m}\left( t,r=a,\vartheta ,\phi \right) =0,\;\text{and
similarly }\Psi _{\omega \ell m}\left( t,r=b,\vartheta ,\phi \right) =0\;.
\end{equation}%
The eigenfunction that satisfy the boundary conditions is%
\begin{eqnarray}
\Psi _{\omega \ell m}\left( t,r,\vartheta ,\phi \right)
&=&c_{0}\dsum\limits_{\left\{ \lambda \right\} }\,e^{-i\omega t}\,r^{-\frac{%
\left( D-2\right) }{2}}\left[ J_{\nu }(\omega r)\,-\frac{J_{\nu }(\omega a)}{%
\,\,N_{\nu }(\omega a)}\,N_{\nu }(\omega r)\right] \,  \notag \\
&&  \notag \\
&&\left\{ \dprod\limits_{\mu =1}^{D-3}\ \sin ^{M_{D-\mu -2}}\theta _{\mu }\
C_{M_{D-\mu -1}-M_{D-\mu -2}}^{M_{D-\mu -2}+\frac{\left( D-\mu -1\right) }{2}%
}\left( \cos \theta _{\mu }\right) \right\} Y_{\ell m}\left( \theta
_{D-2},\phi \right) \,.  \notag \\
&&
\end{eqnarray}%
Where $c_{0}$ is the normalization constant and $\omega $ is the root of the
following transcendental equation

\begin{equation}
J_{\nu }(\omega b)\,\,N_{\nu }(\omega a)-J_{\nu }(\omega a)\,\,N_{\nu
}(\omega b)=0\;\ \text{where\ }\nu =k+\frac{\left( D-2\right) }{2}.
\end{equation}

\bigskip

We define the Casimir energy between two concentric $D$-dimensional spheres
for a massless scalar field

\begin{eqnarray}
E_{C} &=&\frac{1}{2}\sum_{\left\{ \lambda \right\} }\;\omega _{\lambda }\;, 
\notag \\
&&  \notag \\
&=&\frac{1}{2}\sum_{k=0}^{\infty }\left\{ \dprod\limits_{\mu
=1}^{D-3}\dsum\limits_{M_{\mu }=0}^{M_{\mu +1}}\right\}
\dsum\limits_{m=-\ell }^{\ell }\sum_{n=1}^{\infty }\;\omega _{nk},  \notag \\
&=&\frac{1}{2}\sum_{k=0}^{\infty }~g^{\left( D\right) }\left( \nu \right) \
\ \sum_{n=1}^{\infty }\;\omega _{nk}\ \ \ \ \ \ \ \   \notag \\
&&
\end{eqnarray}%
Where $\omega _{n\ell }$ are eigenfrequencies stems from root of the
transcendental equation given in Eq. (9) and the degenaracy of each
eigenfrequency $g^{\left( D\right) }\left( \nu \right) $ $\left( D\text{
represents space dimension}\right) $ could be written as

\begin{equation}
g^{\left( D\right) }\left( \nu \right) =2\nu \frac{\left[ \nu +\frac{1}{2}%
\left( D-4\right) \right] !}{\left( D-2\right) !\left[ \nu -\frac{1}{2}%
\left( D-2\right) \right] !}\ \ \ \text{and\ \ }\nu =k+\frac{\left(
D-2\right) }{2}.
\end{equation}

The zeros of the frequency equation are real and simple \ since $\nu $ is
real and $a$ and $b$ positive. We know that Bessel's series equation are
convergent for all values of the argument. When $\left\vert \omega
\right\vert $ is very large the convergence is so slow. To render the series
useless for the frequency calculation we need the rapidly convergent
evaluation of the Bessel's function formula. A very rapidly convergent
evaluation of the frequency equation can be obtained by using the uniform
asymptotic expansions for the Bessel function. Hence, we should examine the
behavior of the eigenfrequency spectrum for large arguments at fixed $k$ and
large order as $k\rightarrow \infty .$ Thus, to carry out the summation with
respect to $k$ in $E_{C}$, $\ $the sum $\sum_{n=1}^{\infty }\;\omega _{nk}$
\ given in equation (10) replaced by $\sum_{n=1}^{\infty }\;\overline{\omega 
}_{nk}+\sum_{n=1}^{\infty }\;\widetilde{\omega }_{nk}$ where $\widetilde{%
\omega }_{nk}$ \ is the eigenvalue spectrum of the limit $\omega \rightarrow
\infty $ at fixed $k~$\cite{Boyer, Mozcan1,MPRD}. Then, the Casimir energy
which is defined by the eigenfrequency spectrum for large arguments at fixed 
$k$ and large order as $k\rightarrow \infty $ and can be written as

\begin{equation}
E_{C}=\frac{1}{2}\sum_{k=0}^{\infty }~g^{\left( D\right) }\left( \nu \right)
\ \left( \sum_{n=1}^{\infty }\;\overline{\omega }_{nk}+\sum_{n=1}^{\infty }%
\widetilde{\omega }_{nk}\right) .
\end{equation}%
Now, we calculate the eigenfrequencies for large arguments at fixed $\nu .$
We employed the Hankel's asymptotic expansion \cite{Mozcan1, MPRD,Abramowitz}
when $\nu $ is fixed, $\omega a\gg 1$ and $\omega b\gg 1$, one obtains

\begin{equation}
J_{\nu }\left( \widetilde{\omega }a\right) \simeq \sqrt{\frac{2}{\pi 
\widetilde{\omega }a}}\left[ \cos \left( \widetilde{\omega }a-\frac{\nu }{2}%
\pi -\frac{\pi }{4}\right) -\frac{\left( 4\nu ^{2}-1\right) }{8\widetilde{%
\omega }a}\sin \left( \widetilde{\omega }a-\frac{\nu }{2}\pi -\frac{\pi }{4}%
\right) \right]
\end{equation}

\bigskip 
\begin{equation}
N_{\nu }\left( \widetilde{\omega }a\right) \simeq \sqrt{\frac{2}{\pi 
\widetilde{\omega }a}}\left[ \cos \left( \widetilde{\omega }a-\frac{\nu }{2}%
\pi -\frac{\pi }{4}\right) +\frac{\left( 4\nu ^{2}-1\right) }{8\widetilde{%
\omega }a}\sin \left( \widetilde{\omega }a-\frac{\nu }{2}\pi -\frac{\pi }{4}%
\right) \right]
\end{equation}%
and similar expressions for $J_{\nu }\left( \widetilde{\omega }b\right) $
and $N_{\nu }\left( \widetilde{\omega }b\right) $ with $a$ interchanged for $%
b$. Putting (13) and (14) in the frequency equation given by (9), we obtain
the zeros of frequency equation are almost evenly spaced for very large
argument at fixed $\nu $.

\begin{equation}
\widetilde{\omega }_{nk}^{2}\simeq \left( \frac{n\pi }{b-a}\right) ^{2}+%
\frac{\nu ^{2}}{ab}\ \ \ \text{where}\ \ \ n=1,2,3,4,5,6,...\ \ .
\end{equation}%
And, the frequency equation $\overline{\omega }_{nk}$ the first sum is given
in Eq. (12) as the uniform asymptotic expansion of the Bessel function can
be written as

\begin{equation}
f_{\nu }(\nu \overline{\omega },\lambda )=J_{\nu }\left( \nu \overline{%
\omega }\right) \;N_{\nu }\left( \nu \overline{\omega }\lambda \right)
-J_{\nu }\left( \nu \overline{\omega }\lambda \right) \;N_{\nu }\left( \nu 
\overline{\omega }\right) 
\end{equation}%
Where $\lambda =\frac{a}{b}\ (b>a)$ and $\nu =k+\frac{\left( D-2\right) }{2}$%
.

Then, the Casimir energy between the closely spaced two concentric $D$%
-dimensional spheres for a massless scalar field which is defined by the
eigenfrequency spectrum for large arguments at fixed $k$ and large order as $%
k\rightarrow \infty $ can be written as \cite{Mozcan1,MPRD}

\begin{eqnarray}
E_{C}^{\left( D\right) } &=&\frac{1}{2}\sum_{k=0}^{\infty }~g^{\left(
D\right) }\left( \nu \right) \sum_{n=1}^{\infty }\;\overline{\omega }_{nk}+%
\frac{1}{2}\sum_{k=0}^{\infty }~g^{\left( D\right) }\left( \nu \right)
\sum_{n=1}^{\infty }\sqrt{\left( \frac{n\pi }{b-a}\right) ^{2}+\frac{\nu ^{2}%
}{ab}}\ \   \notag \\
&&  \notag \\
&=&\overline{E}_{C}^{\left( D\right) }+\widetilde{E}_{C}^{\left( D\right)
}\;.
\end{eqnarray}%
Where $\overline{\omega }_{nk}$ is the root of the frequency equation given
in Eq. (16).

We consider the first sum defined in Eq. (17). This divergent expression can
be rendered finite by the use of a cutoff or convergence factor. Then$\;$we
define the first sum,

\begin{eqnarray}
\overline{E}_{C}^{\left( D\right) } &=&\frac{1}{2}\sum_{k=0}^{\infty
}~g^{\left( D\right) }\left( \nu \right) ~\sum_{n=1}^{\infty }\;\omega
_{nk}\;e^{-\alpha \omega _{nk}}  \notag \\
&&  \notag \\
&=&\frac{1}{2}\sum_{k=0}^{\infty }~~g^{\left( D\right) }\left( \nu \right)
\;S_{k}\;,
\end{eqnarray}%
where $\omega _{nk}$ refers to $\overline{\omega }_{nk}$ and the factor of $%
e^{-\alpha \omega _{nk}}$ plays the role of an exponential regulator which
effectively suppresses the high frequency contributions to the Casimir
energy, and $S_{k}=\sum_{n=1}^{\infty }\;\omega _{nk}\;e^{-\alpha \omega
_{nk}}$\ is generated by the frequency equation (16). To evaluate the sum $%
S_{k}$, we use the integral representation from the Cauchy's theorem \cite%
{Hag, Mozcan1, MPRD,Mazzitelli} that for two functions $f_{k}(z)$ and $\phi
(z)$ analytic within a closed contour C in which $f_{k}(z)$ has isolated
zeros at $x_{1,}x_{2},x_{3,.........,}x_{n}\;,$

\begin{equation}
\frac{1}{2\pi i}\oint_{C}dz\;\phi (z)\;\frac{d}{dz}\ln f_{k}(z)=\sum_{j}\phi
(x_{j})\;.
\end{equation}%
We choose $\phi (z)=z\;e^{-\alpha z}$ where $\alpha $ is a real positive
constant thus leads to

\begin{equation}
\frac{1}{2\pi i}\oint_{C}dz\;e^{-\alpha z}\;z\frac{d}{dz}\ln f_{\ell
}(z)=\sum_{j}z_{j}\;e^{-\alpha z_{j}}\;.
\end{equation}%
Using this result to replace the sum $S_{k}$ by a contour integral, the
first term of the Casimir energy becomes

\begin{equation}
\overline{E}_{C}=\frac{1}{2}\sum_{k=0}^{\infty }~~g^{\left( D\right) }\left(
\nu \right) \;\frac{1}{2\pi i}\oint_{C}dz\;e^{-\alpha z}\;z\frac{d}{dz}\ln
f_{\nu }(\nu z,\lambda )\;,
\end{equation}%
where the frequency function $f_{\nu }(\nu z,\lambda )$ is given Eq. (16).
The contour $C$ encloses all the positive roots of the equation $f_{\nu
}(\nu z,\lambda )=0$. This contour can be conveniently broken into three
parts \cite{Hag, Mozcan1, MPRD, Mazzitelli}. These consist of a circular
segment $C_{\Gamma }$ and two straight line segments $\Gamma _{1}$ and $%
\Gamma _{2}$ forming an angle $\phi $ and $\pi -\phi $ with respect to the
imaginary axis. When the radius $\Gamma $ is fixed, the contour $C_{\Gamma }$
encloses a finite number of roots of the equation $f_{\nu }(\nu z,\lambda )=0
$. Since the sum of these roots is obviously infinite, the radius $\Gamma $
is a regularization parameter, and taking the limit $\Gamma \rightarrow
\infty $ ( when $\alpha >0\;$) means the removal of the regularization, the
contribution of $C_{\Gamma }$ vanishes provided that $\phi \neq 0$. Hence
the exponential regulator in the Cauchy integral plays the role of the
eliminate of the contribution to the circular part of the contour integral.
Taking the contributions along $\Gamma _{1}$ and $\Gamma _{2}$ which are
complex conjugates of each other and rescaling of integration variable, then
Eq. (18) becomes \cite{Nest1, Hag}

\begin{equation}
\overline{E}_{C}^{\left( D\right) }=-\frac{1}{2\pi b}\lim_{\alpha
\rightarrow 0}\sum_{k=0}^{\infty }~g^{\left( D\right) }\left( \nu \right) \;%
\mathbf{\func{Re}}\;e^{-i\phi }\;\int_{0}^{\infty }dy\;e^{-i\alpha
ye^{-i\phi }/b}\;y\frac{d}{dy}\ln f_{\nu }(\nu ye^{-i\phi },\lambda )\;.
\end{equation}%
Now we use the Lommel's expansions\ or the multiplication theorem for the
function of $f_{\nu }(ye^{-i\phi },\lambda )$ and uniform asymptotic
expansion of the modified Bessel functions one obtains \cite%
{Mozcan1,MPRD,Watson}

\begin{eqnarray}
y\;\frac{d}{dy}\ln \;f_{\nu }(\nu ye^{-i\phi },\lambda ) &=&\frac{\left(
1-\lambda ^{2}\right) ^{2}}{12}\left( \nu ye^{-i\phi }\right) ^{2}+\frac{%
\left( 1-\lambda ^{2}\right) ^{3}}{24}\left( \nu ye^{-i\phi }\right) ^{2} 
\notag \\
&&  \notag \\
&&+\frac{\left( 1-\lambda ^{2}\right) ^{4}}{720}\left[ \left( \nu ye^{-i\phi
}\right) ^{2}(-\nu ^{2}+19)-\left( \nu ye^{-i\phi }\right) ^{4}\right] 
\notag \\
&&  \notag \\
&&+\frac{\left( 1-\lambda ^{2}\right) ^{5}}{1440}\left[ -3\left( \nu
ye^{-i\phi }\right) ^{2}(\nu ^{2}-9)-2\left( \nu ye^{-i\phi }\right) ^{4}%
\right]  \notag \\
&&  \notag \\
&&+\frac{\left( 1-\lambda ^{2}\right) ^{6}}{120960}\left[ \left( \nu
ye^{-i\phi }\right) ^{2}(4\nu ^{4}-290\nu ^{2}+1726)+\left( \nu ye^{-i\phi
}\right) ^{4}(8\nu ^{2}-149)\right.  \notag \\
&&  \notag \\
&&+\left. 4\left( \nu ye^{-i\phi }\right) ^{6}\right] +\left[ \text{Terms in
even powers of }\left( \nu ye^{-i\phi }\right) \right] ,
\end{eqnarray}

\bigskip where $\;\left\vert \lambda ^{2}-1\right\vert <1$ and $\nu =k+\frac{%
\left( D-2\right) }{2}$. Inserting Eq. (23) into Eq. (22) and using the
following integral result

\begin{eqnarray}
I(2n) &=&e^{-i\phi }\int_{0}^{\infty }dy\;e^{-i\alpha ye^{-i\phi
}/b}\;\left( ye^{-i\phi }\right) ^{2n}  \notag \\
&&  \notag \\
&=&i\;\left( -\right) ^{n+1}\left( 2n\right) !\left( \frac{b}{\alpha }%
\right) ^{2n+1}\;,\;\;\;\;\text{where }n=0,1,2,3...
\end{eqnarray}%
then Eq. (22) becomes

\begin{eqnarray}
\overline{E}_{C}^{\left( D\right) } &=&-\frac{1}{2\pi b}\;\lim_{\alpha
\rightarrow 0}\sum_{k=0}^{\infty }g^{\left( D\right) }\left( \nu \right) \;%
\mathbf{\func{Re}\;}\left\{ i\;\frac{\left( 1-\lambda ^{2}\right) ^{2}}{6}%
\nu ^{2}\left( \frac{b}{\alpha }\right) ^{3}+i\;\frac{\left( 1-\lambda
^{2}\right) ^{3}}{12}\nu ^{2}\left( \frac{b}{\alpha }\right) ^{3}\right. 
\notag \\
&&  \notag \\
&&+i\;\frac{\left( 1-\lambda ^{2}\right) ^{4}}{720}\left[ 2\nu ^{2}(-\nu
^{2}+19)\left( \frac{b}{\alpha }\right) ^{3}+24\nu ^{4}\left( \frac{b}{%
\alpha }\right) ^{5}\right]  \notag \\
&&  \notag \\
&&+i\;\frac{\left( 1-\lambda ^{2}\right) ^{5}}{720}\left[ -6\nu ^{2}(\nu
^{2}-9)\left( \frac{b}{\alpha }\right) ^{3}+48\nu ^{4}\left( \frac{b}{\alpha 
}\right) ^{5}\right]  \notag \\
&&  \notag \\
&&+i\;\frac{\left( 1-\lambda ^{2}\right) ^{6}}{120960}\left[ \nu ^{2}(8\nu
^{4}-580\nu ^{2}+3452)\left( \frac{b}{\alpha }\right) ^{3}-24\nu ^{4}(8\nu
^{2}-149)\left( \frac{b}{\alpha }\right) ^{5}+2880\nu ^{6}\left( \frac{b}{%
\alpha }\right) ^{7}\right]  \notag \\
&&  \notag \\
&&+\left. \left[ \text{Terms in imaginary number and even powers of }\nu %
\right] \right\} \;.
\end{eqnarray}%
All terms in the above equation have the singular term in the regulator
parameter $\alpha $ and purely imaginary. Taking the real part of the
parenthesis, thus it leaves the zero result.

\begin{equation}
\overline{E}_{C}^{\left( D\right) }=0\;.
\end{equation}%
The meaning of this result is that there is no contribution from a large
order as $k\rightarrow \infty $ modes for the Casimir energy between the
closely spaced two concentric $D-$dimensional spheres.

Thus, $D$ space dimension of the Casimir energy given in Eq. (17) included
high eigenfrequency modes i.e. $\omega \rightarrow \infty $ at fixed $k$ can
be written as%
\begin{equation}
E_{C}^{\left( D\right) }\;=\frac{1}{2}\sum_{k=0}^{\infty }~g^{\left(
D\right) }\left( \nu \right) \sum_{n=1}^{\infty }\sqrt{\left( \frac{n\pi }{%
b-a}\right) ^{2}+\frac{\nu ^{2}}{ab}}
\end{equation}%
Where \ $\nu =k+\frac{\left( D-2\right) }{2}.$ This divergent sum can be
regularized by using the Abel-Plana sum formula which could be given as \cite%
{Mosbook, Mos1}

\begin{eqnarray}
\text{Reg}\left[ \sum_{n=1}^{\infty }f(n)\right] &=&  \notag \\
&&  \notag \\
\dsum\limits_{n=0}^{\infty }f\left( n\right) -\dint\limits_{0}^{\infty
}f(x)dx &=&\frac{1}{2}f(0)+i\dint\limits_{0}^{\infty }\frac{f(it)-f(-it)}{%
e^{2\pi t}-1}dt.
\end{eqnarray}%
Where $f\left( z\right) $ is an analytic function in the right half plane
and Reg refers to the regularized value of the sum. The other useful
Abel-Plana sum formula for the half integer number is

\begin{equation}
\dsum\limits_{n=0}^{\infty }f\left( n+\frac{1}{2}\right)
=\dint\limits_{0}^{\infty }f(x)dx-i\dint\limits_{0}^{\infty }\frac{%
f(it)-f(-it)}{e^{2\pi t}+1}dt.
\end{equation}

We can rewrite the sum given in Eq. (27) using the Abel-Plana sum formula,
which leads to

\begin{eqnarray}
E_{C}^{\left( D\right) }\; &=&\frac{1}{2}\sum_{k=0}^{\infty }~g^{\left(
D\right) }\left( \nu \right) \ \ \text{Reg}\left[ \sum_{n=1}^{\infty }\sqrt{%
\left( \frac{n\pi }{b-a}\right) ^{2}+\frac{\nu ^{2}}{ab}}\right]  \notag \\
&&  \notag \\
&=&-\frac{1}{4\sqrt{ab}}\sum_{k=0}^{\infty }\nu ~g^{\left( D\right) }\left(
\nu \right) -\frac{1}{2\pi \sqrt{ab}}\xi \sum_{k=0}^{\infty }\nu
^{2}~g^{\left( D\right) }\left( \nu \right) \ \dint\limits_{1}^{\infty }%
\left[ y^{2}-1\right] ^{\frac{1}{2}}\frac{dy}{e^{\xi \nu y}-1}\ .  \notag \\
&&
\end{eqnarray}%
where $\xi =\frac{2d}{\sqrt{ab}}$ and $d=b-a$. The first divergent sum in
Eq. (30) can be removed by using the Hurwitz zeta function \cite%
{Elizaldebook}

\begin{eqnarray}
\sum_{k=0}^{\infty }\left( k+\frac{\left( D-2\right) }{2}\right) ^{p}
&=&\zeta \left( -p,\frac{\left( D-2\right) }{2}\right) ,  \notag \\
&=&-\frac{B_{p+1}\left( \frac{\left( D-2\right) }{2}\right) }{p+1}.
\end{eqnarray}%
Where $p=0,1,2,3,..$ and $B_{p+1}\left( \frac{\left( D-2\right) }{2}\right) $
is the Bernoulli polynomial. To regularized the second term in Eq. (30) we
use the half integer Abel-Plana sum formula given in Eq. (29) for $D-$odd
space dimension and the Abel-Plana sum formula given in Eq. (28) for $D-$%
even space dimension, thus the Casimir energy per unit surface area on the
inner sphere for the specialized cases where our space has dimensions $%
D=4,5,6,7,8,9,10,11$ ( for $D=3$ result is given by in \cite{MPRD} ) can be
written as $\left( \text{the total spherical surface area of }D\text{
dimension is }A^{^{\left( D\right) }}\text{ }=2\pi ^{D/2}\frac{1}{\Gamma
\left( \frac{D}{2}\right) }a^{D-1}\right) $

\bigskip 
\begin{eqnarray*}
\frac{E_{C}^{\left( 4\right) }}{A^{^{\left( 4\right) }}} &=&-\frac{3}{128\pi
^{2}}\left( \frac{\sqrt{ab}}{a}\right) ^{3}\frac{\zeta \left( 5\right) }{%
d^{4}}, \\
&& \\
\frac{E_{C}^{\left( 5\right) }}{A^{^{\left( 5\right) }}} &=&-\frac{1}{32\pi
^{3}}\left( \frac{\sqrt{ab}}{a}\right) ^{4}\frac{\zeta \left( 6\right) }{%
d^{5}}\left[ 1-\frac{1}{8}\eta ^{2}\frac{\zeta \left( 4\right) }{\zeta
\left( 6\right) }\right. \\
&& \\
&&\left. \ \ \ \ \ \ \ \ \ \ \ \ \ \ \ \ \ \ \ \ \ \ \ \ \ \ \ \ \ \ \ \ \ -%
\frac{7}{64}\eta ^{4}\frac{\zeta \left( 4\right) }{\zeta \left( 6\right) }-%
\frac{1}{96}\eta ^{4}\frac{\zeta \left( 2\right) }{\zeta \left( 6\right) }%
\right] , \\
&& \\
\frac{E_{C}^{\left( 6\right) }}{A^{^{\left( 6\right) }}} &=&-\frac{15}{%
1024\pi ^{3}}\left( \frac{\sqrt{ab}}{a}\right) ^{5}\frac{\zeta \left(
7\right) }{d^{6}}\left[ 1-\frac{4}{15}\eta ^{2}\frac{\zeta \left( 5\right) }{%
\zeta \left( 7\right) }\right] ,
\end{eqnarray*}

\begin{eqnarray*}
\frac{E_{C}^{\left( 7\right) }}{A^{^{\left( 7\right) }}} &=&-\frac{3}{128\pi
^{4}}\left( \frac{\sqrt{ab}}{a}\right) ^{6}\frac{\zeta \left( 8\right) }{%
d^{7}}\left[ 1-\frac{5}{12}\eta ^{2}\frac{\zeta \left( 6\right) }{\zeta
\left( 8\right) }+\frac{3}{64}\eta ^{4}\frac{\zeta \left( 4\right) }{\zeta
\left( 8\right) }\right. \\
&& \\
&&\left. +\frac{155}{1536\pi }\eta ^{6}\frac{\zeta \left( 6\right) }{\zeta
\left( 8\right) }+\frac{35}{768\pi ^{2}}\eta ^{6}\frac{\zeta \left( 4\right) 
}{\zeta \left( 8\right) }+\frac{1}{256}\eta ^{6}\frac{\zeta \left( 2\right) 
}{\zeta \left( 8\right) }\right] , \\
&& \\
\frac{E_{C}^{\left( 8\right) }}{A^{^{\left( 8\right) }}} &=&-\frac{105}{%
8192\pi ^{4}}\left( \frac{\sqrt{ab}}{a}\right) ^{7}\frac{\zeta \left(
9\right) }{d^{8}}\left[ 1-\frac{4}{7}\eta ^{2}\frac{\zeta \left( 7\right) }{%
\zeta \left( 9\right) }+\frac{64}{525}\eta ^{4}\frac{\zeta \left( 5\right) }{%
\zeta \left( 9\right) }\right] , \\
&& \\
\frac{E_{C}^{\left( 9\right) }}{A^{^{\left( 9\right) }}} &=&-\frac{3}{128\pi
^{5}}\left( \frac{\sqrt{ab}}{a}\right) ^{8}\frac{\zeta \left( 10\right) }{%
d^{9}}\left[ 1-\frac{35}{48}\eta ^{2}\frac{\zeta \left( 8\right) }{\zeta
\left( 10\right) }+\frac{259}{1152}\eta ^{4}\frac{\zeta \left( 6\right) }{%
\zeta \left( 10\right) }\right. \\
&& \\
&&-\frac{175}{7168}\eta ^{6}\frac{\zeta \left( 4\right) }{\zeta \left(
10\right) }-\frac{4445}{49152\pi ^{6}}\eta ^{8}\frac{\zeta \left( 8\right) }{%
\zeta \left( 10\right) }-\frac{5425}{73729}\eta ^{8}\frac{\zeta \left(
6\right) }{\zeta \left( 10\right) } \\
&& \\
&&\left. -\frac{27195}{1105920\pi ^{2}}\eta ^{8}\frac{\zeta \left( 4\right) 
}{\zeta \left( 10\right) }-\frac{175}{86016}\eta ^{8}\frac{\zeta \left(
2\right) }{\zeta \left( 10\right) }\right] ,
\end{eqnarray*}

\begin{eqnarray*}
\frac{E_{C}^{\left( 10\right) }}{A^{^{\left( 10\right) }}} &=&-\frac{945}{%
65536\pi ^{5}}\left( \frac{\sqrt{ab}}{a}\right) ^{9}\frac{\zeta \left(
11\right) }{d^{10}}\left[ 1-\frac{8}{9}\eta ^{2}\frac{\zeta \left( 9\right) 
}{\zeta \left( 11\right) }+\frac{16}{45}\eta ^{4}\frac{\zeta \left( 7\right) 
}{\zeta \left( 11\right) }\right. \\
&& \\
&&\left. -\frac{256}{3675}\eta ^{6}\frac{\zeta \left( 5\right) }{\zeta
\left( 11\right) }\right] ,
\end{eqnarray*}

\begin{eqnarray}
\frac{E_{C}^{\left( 11\right) }}{A^{^{\left( 11\right) }}} &=&-\frac{15}{%
512\pi ^{6}}\left( \frac{\sqrt{ab}}{a}\right) ^{10}\frac{\zeta \left(
12\right) }{d^{11}}\left[ 1-\frac{21}{20}\eta ^{2}\frac{\zeta \left(
10\right) }{\zeta \left( 12\right) }+\frac{329}{640}\eta ^{4}\frac{\zeta
\left( 8\right) }{\zeta \left( 12\right) }\right.  \notag \\
&&  \notag \\
\ \ \ \ \ \ \ \ \ \ \ \ \ \ \ \ \ &&-\frac{3229}{23040}\eta ^{6}\frac{\zeta
\left( 6\right) }{\zeta \left( 12\right) }+\frac{245}{16384}\eta ^{8}\frac{%
\zeta \left( 4\right) }{\zeta \left( 12\right) }  \notag \\
&&  \notag \\
&&+\frac{10731}{131072\pi ^{8}}\eta ^{10}\frac{\zeta \left( 10\right) }{%
\zeta \left( 12\right) }+\frac{6223}{65536\pi ^{6}}\eta ^{10}\frac{\zeta
\left( 8\right) }{\zeta \left( 12\right) }  \notag \\
&&  \notag \\
&&+\frac{10199}{196608\pi ^{4}}\eta ^{10}\frac{\zeta \left( 6\right) }{\zeta
\left( 12\right) }  \notag \\
&&  \notag \\
&&\left. +\frac{22603}{1474560\pi ^{2}}\eta ^{10}\frac{\zeta \left( 4\right) 
}{\zeta \left( 12\right) }+\frac{569625}{8388608}\eta ^{10}\frac{\zeta
\left( 2\right) }{\zeta \left( 12\right) }\right] .  \notag \\
&&
\end{eqnarray}

Where $\zeta \left( s\right) $ is the Riemann zeta function and $\eta =\frac{%
d}{\sqrt{ab}}$. Our results are interest at the limiting case which is
narrow slit is defined by $\eta =\frac{d}{\sqrt{ab}}\ll 1$ \cite{Brevik}. We
easily analysis that in the limit $a,b\rightarrow \infty $ and $d\rightarrow
0$ ($\eta \rightarrow 0$ and $\frac{\sqrt{ab}}{a}\rightarrow 1$) which means
that the surfaces between two spheres converted to the parallel plate
geometry. One finds that the leading term of the Casimir energy per unit
area for $D$ dimension can be written as

\begin{equation}
\frac{E_{C}^{\left( D\right) }}{A^{\left( D\right) }}=-\frac{1}{\left( 4\pi
\right) ^{\frac{\left( D+1\right) }{2}}\ d^{D}}\Gamma \left( \frac{\left(
D+1\right) }{2}\right) \zeta \left( D+1\right) .
\end{equation}%
This result is exactly the same as the scalar Casimir energy of the parallel
plates for $D$ dimension \cite{JanAmbjorn}. Thus our approach developed here
has been the satisfactory check. As far as we know this result is obtained
here for the first time.

\section{\protect\bigskip CONCLUSION}

\qquad In this work, we have calculated the Casimir energy between closely
spaced two concentric $D-$dimensional spheres for a massless scalar field
satisfying the Dirichlet boundary conditions. We obtain the numerical
results of the Casimir energy between the closely spaced two concentric
spheres in space dimension $D=4$ up to $D=11.$ Although the sign and
magnitude of the Casimir energy for the spherical shell and cavity change
dramatically with the dimension, the sign of the Casimir energy between
closely spaced two concentric $D-$dimensional spheres does not change. We
observed that all spacetime dimensions give us the negative renormalized
vacuum energy by quantum fluctuations between closely spaced two concentric $%
D-$dimensional spheres. This result produces to the sign of stabilazition of
the Casimir energy. The condition of stability will be satisfied between $%
D-1 $ dimensional surfaces close to each other.

The interesting result of our calculations is that any approximation
technique is not needed for our geometry. All contributions in the Casimir
energy for a massless scalar field comes from the higher frequencies for \
fixed $k$ between two surfaces boundary conditions. $k\rightarrow \infty $
frequency modes contribution in the Casimir energy is zero for arbitrary
width of annular region between closely spaced surfaces. Moreover, we find
that the Casimir energy per unit area for a massless scalar field satisfying
the Dirichlet boundary conditions between closely spaced two concentric $D-$%
dimensional spheres is the same $D-$dimensional parallel plates in the limit
case \cite{JanAmbjorn}.

\end{document}